\documentclass{JHEP3}
\usepackage{amsmath}
\usepackage{epsfig}

\newcommand{\zz} {s}
\newcommand{\res} {\operatorname{res}}
\newcommand{\al} {\alpha}
\newcommand{\p}{{\cal P}}
\newcommand{\N}{{\cal N}}

\newcommand{\e}{{\epsilon}}

\newcommand{\ka}{{\kappa}}

\newcommand{\cL}{{\cal L}}
\newcommand{\cO}{{\cal O}}
\newcommand{\tr} {\operatorname{tr}}

\newcommand{\bra}[1] {\left<#1\right|}
\newcommand{\ket}[1] {\left|#1\right>}
\newcommand{\braket}[2] {\left<#1\vphantom{#2}\right|
                         \left.\!\vphantom{#1}{#2}\right>}

\title{Universal regularization for string field theory}

\author{Ehud Fuchs and Michael Kroyter\\
Max-Planck-Institut f\"ur Gravitationsphysik\\
Albert-Einstein-Institut\\
14476 Golm, Germany\\
\email{udif@aei.mpg.de, mikroyt@aei.mpg.de}
}

\abstract{
We find an analytical regularization for string field theory calculations.
This regularization has a simple geometric meaning on the worldsheet,
and is therefore universal as level truncation.
However, our regularization has the added advantage of being analytical.
We illustrate how to apply our regularization to both the discrete and
continuous basis for the scalar field and for the bosonized ghost
field, both for numerical and analytical calculations.
We reexamine the inner products of wedge states, which are known to
differ from unity in the oscillator representation
in contrast to the expectation from level truncation.
These inner products describe also the descent relations of string vertices.
The results of applying our regularization strongly suggest that these inner
products indeed equal unity.
We also revisit Schnabl's algebra
and show that the unwanted constant vanishes when using our regularization
even in the oscillator representation.
}

\preprint{{\tt hep-th/0610298}\\AEI-2006-080}

\begin{document}

\section{Introduction}

Regularization should be made with care for issues such as symmetries,
analyticity and the origin of divergences. One amusing example is
given by regularizing the bosonic and fermionic contributions
to the vacuum energy in the context of the DBI
action~\cite{Fradkin:1985qd,Metsaev:1987qp}.
In both cases one gets an infinite set of constant contributions, but
they should be regularized differently since they originate from
integer and half-integer modes respectively.
For the bosonic case one gets
\begin{equation}
        \sum_{n=1}^\infty 1\equiv \zeta(0)=-\frac{1}{2}\,,
\end{equation}
while for the fermionic case
\begin{equation}
        \sum_{n=\frac{1}{2}}^\infty 1\equiv \zeta(0,\frac{1}{2})=0\,.
\end{equation}

In this paper we address the issue of regularization in
bosonic string field theory~\cite{Witten:1986cc}.
Here, divergences can occur due to the singular nature of the
string vertices that act as delta functionals. Witten addressed these
divergences by defining the delta functionals using strips with finite width
in the limit where this width tends to zero.
Witten's construction was made more explicit using
oscillators~\cite{Gross:1987ia,Gross:1987fk,Samuel:1986wp,Ohta:1986wn,Cremmer:1986if}
and CFT
methods~\cite{LeClair:1989sp,LeClair:1989sj}\footnote{A
variation of the oscillator construction is given by
Moyal string field theory~\cite{Bars:2001ag,Bars:2002nu,Bars:2002yj}.}.
All these methods give expressions for Witten's vertex with strips of
zero width.

Actually, it turns out that usually no regularization is required
as long as we do not care about the overall normalization.
In cubic string field theory only the two- and three-vertices are needed,
and since the two-vertex has trivial normalization,
any non-trivial normalization of the three-vertex can be absorbed
into field redefinitions.

Yet, this is not always the case, since in practice, other vertices
are also used. The most prominent example is Schnabl's analytic
solution to the equation of motion~\cite{Schnabl:2005gv}.
In this work, Schnabl constructed the solution
using a novel gauge
and proved Sen's first
conjecture~\cite{Sen:1999mh,Sen:1999xm}.
Following his construction the consistency of the solution was
verified~\cite{Okawa:2006vm,Fuchs:2006hw} and Sen's third conjecture
was proven~\cite{Ellwood:2006ba}.
Generalizations of his construction for the toy model introduced
in~\cite{Gaiotto:2002uk} and exploration of their geometric meaning
were presented in~\cite{Rastelli:2006ap}.
A continuous basis oscillator representation
for Schnabl's operators was found in~\cite{Fuchs:2006an} and
a study of scattering amplitudes
in a variation of his gauge was performed in~\cite{Fuji:2006me}.

Schnabl's construction uses an infinite sum of wedge states.
Wedge states are surface states that are naturally related to string
vertices~\cite{Rastelli:2000iu,Furuuchi:2001df,Rastelli:2001vb,Schnabl:2002gg}.
For integer $n$ the wedge state $\ket{n}$ can be defined as the contraction
of the $n$ string field vertex $\ket{V_n}$ with $n-1$ vacua.
While Schnabl's universal construction is safe from
normalization anomalies, they can occur when working with an
explicit CFT. Indeed, when
choosing to work with 26 scalar fields plus the bosonized ghost, all represented
in the continuous basis, an anomaly occurs in the algebra of Schnabl's
operators~\cite{Fuchs:2006an}. This should be interpreted as an artifact, stating
that the regularization procedure used is anomalous.

Another, even more called for example of such an anomaly comes from evaluating
the descent relation among string field vertices
$\braket{V_1}{V_3}=\ket{V_2}$ or the analogous
equation with wedge states $\braket{1}{3}=1$. One gets the second equation
from the first upon contraction with two vacuum states with ghost numbers
zero and three. The first evaluation of this expression in the oscillator
representation was attempted in the continuous
basis~\cite{Belov:2002pd,Fuchs:2002wk,Belov:2002sq,Fuchs:2003wu,Belov:2003qt}\footnote{The
source of singularities as seen in the continuous basis is that the matrices
defining the wedge states are diagonal and are therefore proportional to the
delta function. Even worse is the behaviour of the Virasoro operators,
which are proportional to complex delta functions in the
continuous basis~\cite{Douglas:2002jm,Fuchs:2002wk,Belov:2002te}.}.
There, it was found that the result of the inner product is not unity.
In order to identify the source of this problem, numerical evaluations of these
expressions using oscillator level truncation in the discrete basis were
performed in~\cite{Fuchs:2005ej,Aref'eva:2006pu}. These papers illustrated that
the problem is not related to the use of the continuous basis.

In the seminal paper of LeClair, Peskin and Preitschopf~\cite{LeClair:1989sj}
a formal justification for the descent relations was given.
It relied on the fact that the relation
\begin{equation}
\label{VirasoroNorm}
\bra{0}\exp\Big({\sum_{n=0}^\infty v_n L_n}\Big)
    \exp\Big({\sum_{n=0}^\infty u_n L_{-n}}\Big)\ket{0}=1\,,
\end{equation}
holds as long as $v_n, u_n$ are small enough and the total central
charge is zero.
One criterion to check whether $v_n, u_n$ are indeed small enough is to turn on
a central charge. If the above correlator diverges, it could be
a sign of trouble.
Unfortunately, it diverges in the case of the vertices.
This raises the question of whether we can trust the descent relations
\begin{equation}
\label{DescentRelations}
\braket{V_1}{V_{N}}=\ket{V_{N-1}},
\end{equation}
or is it necessary to add some normalization factors to these relations.
In this paper we find an analytical regularization under which these
relations hold.

The apparent validity of~(\ref{VirasoroNorm}) is a direct consequence
of the Virasoro algebra with zero central charge.
This suggest that any universal regularization should work.
Universality refers here to independence of the choice of CFT, as long as
the total system has a vanishing central charge.

The regularization we suggest is to shrink the string field states using
the conformal map $f_\zz(\xi)=\zz\xi$, where $\xi$ is a coordinate on the
upper half-plane with the half unit circle as the local coordinate patch
and $\zz<1$ is the dilatation parameter.
Geometrically this regularization can be viewed
as reinstating Witten's prescription of using a strip
with half-width of $\log \zz^{-1}$,
as can be seen in the $\tau,\sigma$ coordinates.
This regularization respects
the symmetry of the problem since it is
universal by construction while resolving the origin of
the singularity,
i.e. the delta-functional gluing and the conical singularity of the
string vertices, see figure~\ref{fig:Vertices}.
\FIGURE{
\label{fig:Vertices}
\epsfig{figure=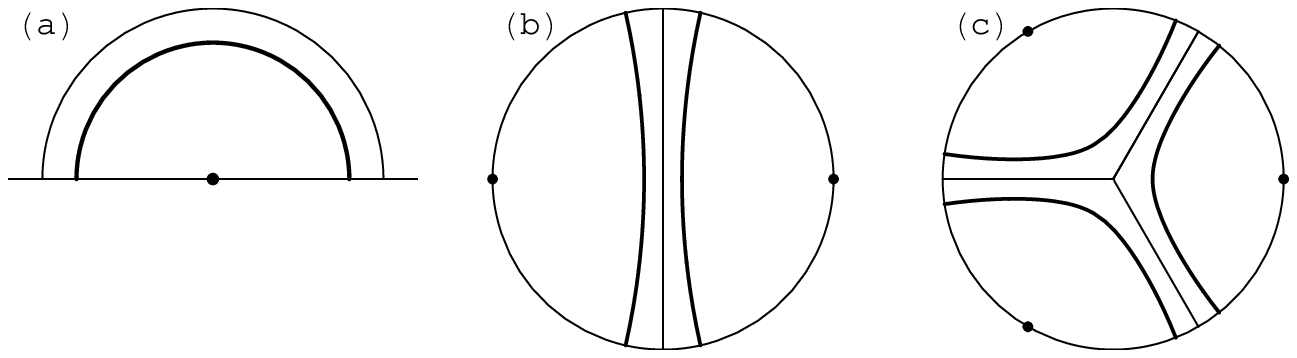}
\caption{The effect of the regularization on the canonical
coordinate patch~(a), the two-vertex~(b) and the three-vertex~(c).}
}

Expanding a field $\Psi$ in level components $\Psi_h$,
the regularized field can be expressed as
\begin{equation}
\Psi^\zz \equiv f_\zz\circ\Psi=\zz^{L_0}\Psi=\sum_h \zz^h \Psi_h\,.
\end{equation}
This shows that our regularization is closely related to level
truncation~\cite{Kostelecky:1990nt}.
Specifically, writing
\begin{equation}
\braket{\Psi_1^\zz}{\Psi_2^\zz}=g_{12}(\zz)\,,
\end{equation}
one can obtain the level $(L,2L)$ result from Taylor expanding $g_{12}(\zz)$
to order $2L$ and then setting $\zz=1$.
This also demonstrates the down side of level truncation.
The function $g_{12}(\zz)$ can have a well defined Taylor expansion, but with
radius of convergence less than unity. Calculations to any finite level
will not yield a divergent result.

Our regularization has the added feature of analyticity, which is missing
in level truncation.
To demonstrate this we can look at the overlap of two butterfly states
\begin{equation}
\label{Butterfly}
\braket{B_\alpha}{B_\beta} =
    \bra{0}\exp\Big(\frac{1}{2}\alpha L_2\Big)
    \exp\Big(\frac{1}{2}\beta L_{-2}\Big)\ket{0} =
    (1-\alpha\beta)^{-\frac{c}{8}} .
\end{equation}
This case is simple enough to derive the explicit normalization.
Setting $\alpha=\beta$ we can calculate the normalization of the butterfly
state.
It is clear that for $|\alpha|>1$ the normalization is not well defined.
Still, truncating to any finite level gives a finite result.
Calculating the normalization in our regularization simply amounts
to rescaling $\alpha\rightarrow \zz^2\alpha$.
Notice, that even though the level truncation calculation simply amounts
to repeatedly applying the Virasoro algebra, it is quicker to use
our regularization and then Taylor expand the result.
The limiting case $\alpha=1$ is the interesting case where the butterfly
becomes a projector. The normalization is only well defined if we apply
our regularization. Geometrically, in this case the boundary of the
surface exactly pinches the coordinate patch (at the mid-point).
Our regularization shrinks the state to avoid this singularity.

The effect of our regularization on squeezed states is simply to
rescale its defining matrix and vector,
\begin{equation}
\label{Squeezed}
\ket{S^\zz}=\exp(a^\dagger S^\zz a^\dagger + V^\zz a^\dagger)\ket{0},\qquad
S^\zz_{nm} = S_{nm}\zz^{n+m}\,,\quad
V^\zz_n = V_n \zz^n\,.
\end{equation}
This means that we can still use squeezed state techniques in our calculations,
which are usually more efficient than level truncation calculations.
Actually, many calculations in oscillators were done by truncating
the matrix $S_{nm}$ to finite size. This
method, named oscillator level truncation,
can be used to harness the power of squeezed state techniques.
In particular it was used in~\cite{Rastelli:2001hh} for defining
the spectral density of the continuous basis and
in~\cite{Taylor:2002bq,Taylor:2004rh} for calculating scattering
amplitudes.
However, as stated above, oscillator level truncation can give anomalous results.
Normalization calculations in the continuous basis also suffer
from the same inconsistencies. In this paper we show how to apply
our regularization to the discrete basis and to the
continuous basis and demonstrate that it gives consistent results
in both cases.

To calculate the normalization of the descent
relations~(\ref{DescentRelations}) we contract them with $N-1$ vacuum
states. This leaves us with two wedge states
\begin{equation}
\braket{1}{N} \equiv \gamma_{1,N}\,.
\end{equation}
Using~(\ref{VirasoroNorm}) we get $\gamma_{1,N}=1$, but as
we demonstrated with the butterfly calculation~(\ref{Butterfly}),
one must use~(\ref{VirasoroNorm}) with caution.

In this paper we analyze
the more general wedge state inner product
$\gamma_{n_1,n_2}\equiv\braket{n_1}{n_2}$
using our regularization.
We show that
$\gamma_{2+t_1,2+t_2}=1+\cO(t_1^3 t_2^3)$.
We also prove that 
(the expression with the most potential for singularities)
$\gamma_{1,1}=1$,
and evaluate
$\gamma_{1,3}$ numerically showing that it is consistent with
being unity for values of $\zz$ at least up to $\zz=0.999$.
This seems sufficient to claim that
$\gamma_{n_1,n_2}=1$ in the context of our regularization.
We believe that the contraction of any two legitimate surface
states, i.e. states whose defining surface contains the local coordinate patch,
would give unity, when using our regularization.

Our regularization is for states.
It can be generalized to operators, but only if one is careful to
use it in the right context.
The ambiguity comes from the fact that an operator of weight $h$
can be scaled as $\zz^h$ or $\zz^{-h}$ depending on whether it appears in
a ket or a bra state.
To avoid this ambiguity we limit the use of regularized operators
to computations involving normal ordered states.
Then creation operators will always appear in ket states while
annihilation operators will always appear in bra states.
There is a subtlety with the ghost zero-modes,
which is resolved by acting on either of the vacua with the $c_1, c_0, c_{-1}$
ghost operators. This is needed to saturate the ghost number anomaly.
Non unitary CFT's, like the scalar field with time signature, are another
issue. We will postpone the question of regularizing operators
like $:e^{ikX}:$ to a later time.
For now, we present a few examples.
We start with the Virasoro operators.
Under the above conditions the regularized Virasoro operators are
\begin{equation}
L_n^\zz = \zz^{|n|}L_n\,.
\end{equation}
These operators satisfy the regularized Virasoro algebra
\begin{equation}
\label{VirasoroAlgebra}
[L_n^\zz, L_m^\zz] = \zz^{|n|+|m|-|n+m|}
    \left( (n-m)L_{n+m}^\zz + \frac{c}{12}(n^3-n)\delta_{n+m} \right).
\end{equation}
This algebra is guaranteed to satisfy the Jacobi identity by construction.
For the scalar field we have (ignoring zero-modes)
\begin{equation}
\label{ScalarAlgebra}
\alpha_n^\zz = \zz^{|n|}\alpha_n\,,\qquad
[\alpha_n^\zz,\alpha_m^\zz] = \zz^{|2n|}n\delta_{n+m}\,.
\end{equation}
In terms of these operators the Virasoro operators are
\begin{align}
\qquad L_n^\zz &= \frac{1}{2}\sum_{m=1}^{n-1}\alpha_m^\zz \alpha_{n-m}^\zz +
    \sum_{m=n+1}^{\infty} \zz^{2(n-m)}\alpha_m^\zz\alpha_{n-m}^\zz
    & n>0\,, \qquad \\
\qquad L_n^\zz &= \frac{1}{2}\sum_{m=n+1}^{-1}\alpha_m^\zz \alpha_{n-m}^\zz +
    \sum_{m=-\infty}^{n-1} \zz^{2(m-n)}\alpha_m^\zz\alpha_{n-m}^\zz
    & n<0\,. \qquad
\end{align}
It might seem that these expressions diverge since we have negative
powers of $\zz<1$, but one can check that in the dependence
of $L_n^\zz$ on the original oscillators there are only positive
powers of $\zz$.

The rest of the paper is structured as follows.
In section~\ref{sec:NewReg} we show how to regularize continuous
basis calculations.
In section~\ref{sec:Schnabl}
we analyze Schnabl's algebra using our regularization and
generalizations thereof.
In section~\ref{sec:V1V2V3} we calculate the descent relations
in the continuous and in the discrete basis.
We conclude in section~\ref{sec:conc}.

\section{The continuous basis}
\label{sec:NewReg}

Our regularization is analytic with respect to $\zz$,
unlike level truncation regularization, where the
dependence on the level is encoded in a step-function.
From~(\ref{Squeezed}), we see that for calculations
involving squeezed states
we should add, between every two contracted indices,
the regularizing matrix
\begin{equation}
\label{OurReg}
\rho_\zz^{nm} \equiv \delta_{n,m}\zz^{2n}\,.
\end{equation}
However, we would also like to work in the continuous basis
to get analytical expressions, whenever possible.
To that end, we evaluate $\rho_\zz$ in the continuous basis
in~\ref{sec:RegEval}. Then, in~\ref{sec:RegUseRules} we illustrate
the properties of the resulting expression.

\subsection{Evaluating the form of the regulator in the continuous basis}
\label{sec:RegEval}

The continuous basis is the basis that diagonalizes
$K_1=L_1+L_{-1}$~\cite{Rastelli:2001hh}.
The oscillators in this basis can be written as a linear
combination of the discrete basis oscillators,
\begin{equation}
a^\dagger_\ka=\sum_{n=1}^\infty
 \frac{v_n^{\ka}}{\sqrt{\N(\ka)}}
  \,a^\dagger_n\,, \qquad
a_\ka=\sum_{n=1}^\infty
\frac{v_n^{\ka}}{\sqrt{\N(\ka)}}\, a_n\,,
\end{equation}
where $-\infty<\ka<\infty$.
The transformation matrix is defined by the generating function,
\begin{equation}
\label{GenFunc}
f_\ka(z)=\frac{1-e^{-\ka \tan^{-1}z}}{\ka}
  \equiv \sum_{n=1}^\infty\frac{v_n^\ka}{\sqrt{n}}z^n\,.
\end{equation}
The normalization,
\begin{align}
\N(\ka)&=\frac{2}{\ka}\sinh(\frac{\ka \pi}{2})\,,
\end{align}
gives orthogonality and completeness relations of the form
\begin{align}
\label{vvDeltanm}
\int_{-\infty}^\infty d\ka \frac{v_n^{\ka}v_m^{\ka}}{\N(\ka)}=&
 \delta_{nm}\,,\\
\label{vvDeltakaka}
\sum_{n=1}^\infty v_n^{\ka}v_n^{\ka'}=&\N(\ka)\delta(\ka-\ka')\,.
\end{align}
We can write the regularizing matrix~(\ref{OurReg}) in the continuous basis
as\footnote{We write $\rho_s$ and $\rho_\e$ interchangeably, keeping~(\ref{eZ}) in mind.}
\begin{align}
\label{rhoDef}
\rho_\e(\ka,\ka') &=
    \frac{1}{\sqrt{\N(\ka)\N(\ka')}}    \sum_{n,m=1}^\infty v_n^\ka v_m^{\ka'}\rho_\zz^{nm} =
    \frac{1}{\sqrt{\N(\ka)\N(\ka')}}
     \sum_{n=1}^\infty v_n^\ka v_n^{\ka'} \zz^{2n}\,,\\
\label{eZ}
\e &\equiv  1-\zz\,.
\end{align}
From the operator regularization point of view this matrix can be
interpreted as the commutation relation
\begin{equation}
[a_\ka^\zz,{a_{\ka'}^\zz}^\dagger] = \rho_\e(\ka,\ka')\,.
\end{equation}
This expression is reminiscent of the expressions that one gets in level
truncation~\cite{Belov:2002pd,Fuchs:2002wk,Belov:2002sq}.
However, here the regularization matrix is non-diagonal and thus
resolves the singular delta-type behaviour.

We can write the regularizing matrix explicitly using
\begin{align}
\sum_{n=1}^\infty v_n^\ka v_n^{\ka'} \zz^{2n}=\frac{\zz}{2}
  \partial_\zz \frac{1}{2\pi i}\oint \frac{du}{u}
      f_\ka(u)f_{\ka'}\big(\frac{\zz^2}{u}\big)\,.
\end{align}
It is easy to see that only the term involving two
exponents gives a non-zero result. Rescaling $u$ we get,
\begin{align}
\sum_{n=1}^\infty v_n^\ka v_n^{\ka'} \zz^{2n}=\frac{\zz}{2}
  \partial_\zz \frac{1}{2\pi i}\oint \frac{du}{u}
   \frac{e^{-\ka\tan^{-1}(\zz u)-\ka'\cot^{-1}(\frac{u}{\zz})}}{\ka\ka'}\,.
\end{align}
We now take $u$ on the unit circle, where the integral is well
defined for all $\zz<1$ and expand the argument around $\zz=1$,
\begin{align}
\sum_{n=1}^\infty v_n^\ka v_n^{\ka'} \zz^{2n}=
-\frac{1-\e}{4\pi \ka \ka'}
 \partial_\e \int_0^{2\pi} d\theta
  e^{-\ka_+ \tan^{-1}\left(\frac{\cos\vphantom{^2} \theta}{\e}
    \left(1-\frac{\e}{2}+\cO(\e^2)\right)\right)
   -i \ka_- \tanh^{-1}\left(\sin \theta(1-\frac{\e^2}{2}) +\cO(\e^2)\right)},
\end{align}
where we define
\begin{align}
\ka_{\pm} \equiv \frac{\ka\pm\ka'}{2}\,.
\end{align}
We know from the derivation of the level-truncation
expression $\rho_{\text{fin}}$
that this expression diverges
as $\log \e$. We are thus, not interested in $\cO(\e)$ corrections
to the expression, because they would not contribute even for a
product of regulators. We also note that the argument of the exponent
is bounded. Thus, we can neglect higher order terms.
Moreover, a rescaling of $\e\rightarrow \e(1+\frac{\e}{2})$
gives,
\begin{align}
\sum_{n=1}^\infty v_n^\ka v_n^{\ka'} \zz^{2n}=
-\frac{1}{4\pi \ka \ka'}
 \partial_\e \int_0^{2\pi} d\theta
  e^{-\ka_+ \tan^{-1}\left(\frac{\cos \theta}{\e}\right)
   -i \ka_- \tanh^{-1}\left(\sin \theta(1-\frac{\e^2}{2})\right)}.
\end{align}
We now use the obvious symmetries and find that up to the
relevant accuracy we have,
\begin{align}
\rho_\e(\ka,\ka')=\frac{I_1+I_2}{\pi \ka \ka'\sqrt{\N(\ka)\N(\ka')}}\,.
\end{align}
Here,
\begin{align}
\nonumber
& I_1=- \int_0^\frac{\pi}{2}
\frac{\ka_- \e\sin \theta d \theta}{\cos^2\theta+\e^2 \sin^2\theta}
  \cosh \big(\ka_+ \tan^{-1}(\frac{\cos\theta}{\e})\big)
  \sin \Big(\ka_- \tanh^{-1}\big((1-\frac{\e^2}{2}) \sin\theta\big)\Big)=\\
&- \int_0^1
\frac{\ka_- \e dx}{x^2+\e^2 (1-x^2)}
  \cosh \big(\ka_+ \tan^{-1}(\frac{x}{\e})\big)
  \sin \Big(\ka_- \tanh^{-1}\big((1-\frac{\e^2}{2}) \sqrt{1-x^2} \,\big)\Big),\\
& I_2=\int_0^\frac{\pi}{2} \frac{\ka_+ \cos\theta  d \theta}
 {\e^2+\cos^2 \theta}
   \sinh \big(\ka_+ \tan^{-1}(\frac{\cos\theta}{\e})\big)
   \cos\Big(\ka_- \tanh^{-1}\big((1-\frac{\e^2}{2}) \sin\theta\big)\Big)=
\nonumber\\
&\int_0^1
\frac{\ka_+ x dx}{(x^2+\e^2) \sqrt{1-x^2}}
  \sinh \big(\ka_+ \tan^{-1}(\frac{x}{\e})\big)
  \cos \Big(\ka_- \tanh^{-1}\big((1-\frac{\e^2}{2}) \sqrt{1-x^2} \,\big)\Big),
\end{align}
where $x=\cos \theta$.

Inspecting the integrands we see that the trigonometric and
hyperbolic functions are bounded, while the terms multiplying them peak
around $x=\e$. This produces the expected logarithmic singularity
for $\ka_- =0$. For $\ka_- \neq 0$, the trigonometric function
tempers the divergence. Thus, we can add terms that are $\cO(\e)$
to the integrands. Up to such terms, we note that the two integrals
are related by an integration by parts with zero boundary terms,
\begin{align}
I_2=-\frac{\ka_+ ^2}{\ka_-}I_1\,.
\end{align}
Next, we use the fact that the integrand of $I_1$ is non-zero only
at a small neighbourhood of $\e$ in order to drop more terms,
rescale to $y=\e^{-1}x$ and continue the $y$ integration to
infinity. This gives us,
\begin{align}
\nonumber
\rho_\e(\ka,\ka')=&\Im \frac{e^{i \ka_- \log\frac{\e}{2}}}
   {\pi \ka_- \sqrt{\N(\ka)\N(\ka')}}
\int_0^\infty \frac{dy}{1+y^2}
e^{i \ka_- \frac{\log(1+y^2)}{2}}
  \cosh \big(\ka_+ \tan^{-1}y\big)\\
\nonumber
  =&\frac{2}{\ka \ka'\sqrt{\N(\ka)\N(\ka')}}
     \Re \frac{e^{i \ka_- \log \e}}
       {B(\frac{i \ka'}{2},-\frac{i \ka}{2})}\\
\label{rho}
  =&\frac{1}{\ka \ka'\sqrt{\N(\ka)\N(\ka')}}
     \left( \frac{e^{i \ka_- \log \e}}
       {B(\frac{i \ka'}{2},-\frac{i \ka}{2})}+
       \frac{e^{-i \ka_- \log \e}}
       {B(\frac{i \ka}{2},-\frac{i \ka'}{2})}\right),
\end{align}
where $B$ is the beta function.

\subsection{Calculating with the regulator in the continuous basis}
\label{sec:RegUseRules}

As a first observation we note that in the limit $\ka_- \rightarrow 0$,
(\ref{rho}) reduces to the known result (note that $2\e$ here equals
$\e$ in~\cite{Fuchs:2002wk}),
\begin{align}
\rho_\e(\ka)=-\frac{2\log \e+2\gamma+
   \psi(\frac{i \ka}{2})+\psi(-\frac{i \ka}{2})}{4\pi}\,.
\end{align}
However, we should not impose this limit.
Rather, $\rho_\e(\ka,\ka')$ should behave as $\delta(\ka-\ka')$
in the limit $\e\rightarrow 0$,
as is demonstrated below.

The regulator $\rho_\e$ is a matrix. Matrix multiplication in the
continuous basis is integration. Given a vector $V$, that is
a function of $\ka$, we can evaluate the product $\rho_\e V$.
If $V$ has no real singularities and does not diverge
faster than $e^{\frac{|\ka|\pi}{4}}$ for $\ka\rightarrow \pm \infty$
we have
\begin{align}
\label{rhoVV}
(\rho_\e V)(\ka)=V(\ka)+\cO(\e)\,.
\end{align}
To show this,
we note that for a function that does not diverge too
fast for imaginary argument, we can close the contour of integration
neglecting the contribution of the arc, so that the integral is given
by a sum over residues. This is clear for a function that does not diverge
super-exponentially fast in the imaginary directions (other than the
singularities that we would have to sum) and is also true for a
function that diverges faster, as can be seen by changing simultaneously
$\e$ and the radius of integration ($e^{-\ka^2}$ is an example for such a
function). Note that we have two integrands, one of which should be
closed in the upper half plane and the other in the lower half plane.
However, we cannot separate these integrands, since they have poles for
$\ka'=\ka$ that cancel for their sum. Thus, we have to first deform
the integration contour, say by lowering it by $\delta$ below the real axis
in the complex $\ka'$ plane. Then, we can close the contours separately.
The integrand proportional to $e^{i \ka_- \log \e}$ should be closed
from below and picks up residues at $\ka'=\ka-2i n$, as well as at
poles of $V$ in the lower half plane. The second integrand picks up
similar residues at the upper half plane, as well as the residue at
$\ka'=\ka$. Generically, we could also
have cuts coming from the square root terminating at $\ka'=\pm 2i n$
as well as (possibly) other cuts coming from $V$.
These cuts can be taken into account by integrating around them.
The result of the summation of residues and cut integrals
would be a function of $\e$, which has
a positive radius of convergence. It is possible to change the order
of summation and the limit for this function. Then, all the contributions
that come from either the upper or the lower half plane would vanish
in the $\e\rightarrow 0$ limit. The poles at $\ka'=\pm 2i n$ for example
would all be proportional to $\e^{2n}$. Thus, we arrive at the important
observation, that the only term we should evaluate is the residue
at $\ka'=\ka$ of the integrand proportional to
$e^{-i \ka_- \log \e}$. Direct evaluation then
proves~(\ref{rhoVV})\footnote{We could of course also deform the contour
by a positive amount in the imaginary direction, such that only the other
term would contribute. The result is of course the same.}.
We will henceforth omit the $\cO(\e^k)$. All identities should be
understood as holding up to such negligible terms.

We continue by evaluating the matrix multiplication
\begin{align}
(\rho_\e\rho_{\e'})(\ka,\ka')=&
\frac{1}{\ka \ka'\sqrt{\N(\ka)\N(\ka')}}\,\cdot\\
\nonumber\cdot &
 \int_{-\infty}^\infty \frac{d\tilde \ka}{\tilde \ka^2 \N(\tilde \ka)}
     \Big( \frac{e^{i \frac{\ka-\tilde\ka}{2} \log \e}}
       {B(\frac{i \tilde\ka}{2},-\frac{i \ka}{2})}+
       \frac{e^{-i \frac{\ka-\tilde\ka}{2} \log \e}}
       {B(\frac{i \ka}{2},-\frac{i \tilde\ka}{2})}\Big)
          \Big( \frac{e^{i \frac{\tilde\ka-\ka'}{2} \log \e'}}
       {B(\frac{i \ka'}{2},-\frac{i \tilde\ka}{2})}+
       \frac{e^{-i \frac{\tilde\ka-\ka'}{2} \log \e'}}
       {B(\frac{i \tilde\ka}{2},-\frac{i \ka'}{2})}\Big).
\end{align}
Again we should shift the integration contour slightly below the
real axis in order to evaluate the terms separately.
The term that we get from multiplying the second summand in the
first parentheses by the first summand in the second ones is
proportional to $e^{i \frac{\tilde \ka}{2} \log (\e\e')}$.
This contour should be closed from below and would therefore
not contribute in the limit from similar
considerations to those described above.
For the term proportional to $e^{-i \frac{\tilde \ka}{2} \log (\e\e')}$
we have to sum the residues on the real axis. This gives
\begin{align}
2\pi i(\res_{\tilde\ka=\ka}+\res_{\tilde\ka=\ka'})=
\frac{e^{i \frac{\ka-\ka'}{2} \log \e}}
       {B(\frac{i \ka'}{2},-\frac{i \ka}{2})}+
\frac{e^{-i \frac{\ka-\ka'}{2} \log \e'}}
       {B(\frac{i \ka}{2},-\frac{i \ka'}{2})}\,.
\end{align}
The other two terms are proportional to
$e^{\pm i \frac{\tilde \ka}{2} \log (\frac{\e}{\e'})}$.
Since generically $\pm\log (\frac{\e}{\e'})$ is not small,
we have to evaluate the integral exactly.
Assuming without loss of generality that $\e'<\e$ we get for the
$e^{-i \frac{\tilde \ka}{2} \log (\frac{\e}{\e'})}$
term, residues at $\tilde\ka=\ka'-2n i$, which we sum to get
\begin{align}
-2\pi i\sum_{\res}=
\frac{e^{i \frac{\ka-\ka'}{2} \log (\e+\e')}-
    e^{i \frac{\ka-\ka'}{2} \log \e}}
  {B(\frac{i \ka'}{2},-\frac{i \ka}{2})}\,.
\end{align}
For the last term we have to consider the poles at $\tilde\ka=\ka'+2n i$,
which give a similar expression
\begin{align}
2\pi i\sum_{\res}=
\frac{e^{-i \frac{\ka-\ka'}{2} \log (\e+\e')}-
    e^{-i \frac{\ka-\ka'}{2} \log \e}}
  {B(-\frac{i \ka'}{2},\frac{i \ka}{2})}\,,
\end{align}
as well as the poles at $\tilde\ka=\ka,\ka'$, which give
\begin{align}
2\pi i(\res_{\tilde\ka=\ka}+\res_{\tilde\ka=\ka'})=
\frac{e^{-i \frac{\ka-\ka'}{2} \log \e}-
      e^{-i \frac{\ka-\ka'}{2} \log \e'}}
   {B(-\frac{i \ka'}{2},\frac{i \ka}{2})}\,.
\end{align}
All in all we are left with the elegant expression
\begin{align}
\label{rhorho}
(\rho_\e\rho_{\e'})(\ka,\ka')=
\frac{1}{\ka \ka'\sqrt{\N(\ka)\N(\ka')}}
  \Big(\frac{e^{i \frac{\ka-\ka'}{2} \log (\e+\e')}}
  {B(\frac{i \ka'}{2},-\frac{i \ka}{2})}+
  \frac{e^{-i \frac{\ka-\ka'}{2} \log (\e+\e')}}
    {B(-\frac{i \ka'}{2},\frac{i \ka}{2})}\Big)=
\rho_{\e+\e'}(\ka,\ka')\,.
\end{align}

We should however keep in mind, that we cannot iterate this
expression without a limit, since we have used the fact that $\e$ is
small, both in deriving the form of $\rho_\e$~(\ref{rho}) and
in the evaluation above. For example, from its definition it is
clear that $\e<1$, while iterating~(\ref{rhorho}) can lead to $\e>1$.
In fact, in the discrete basis this expression is trivial, and simply
amounts to
\begin{align}
\rho_s \rho_{s'}=\rho_{s s'}\,.
\end{align}
For small $\e$
this gives~(\ref{rhorho}).

\section{Regularizing Schnabl's algebra}
\label{sec:Schnabl}

In this section we address the algebra of Schnabl's operators using our regularization.
In~\ref{sec:SchnablOs} we prove analytically that, when regulated,
these operators obey the correct algebra also in their continuous basis
representation. Next, in~\ref{sec:SchnablVir} we calculate the subleading
terms of the algebra with regularized operators in the Virasoro basis.
Then, in~\ref{sec:SchnablAlt} we study possible generalizations of our
regularization.

\subsection{Using oscillators}
\label{sec:SchnablOs}

In~\cite{Fuchs:2006an} we noted that Schnabl's operators $\cL_0,\cL_0^\dag$
have a very nice structure in the continuous basis,
\begin{align}
\cL_0 &= \int_{-\infty}^\infty d\ka d\ka' \Big(
    \frac{1}{2} a_\ka A(\ka,\ka') a_{\ka'} +
    a_\ka^\dagger B(\ka,\ka') a_{\ka'} \Big) +
    \int_{-\infty}^\infty d\ka \; V_1(\ka) a_\ka + \frac{1}{2}V_{00}\,,\\
\cL_0^\dagger &= \int_{-\infty}^\infty d\ka d\ka' \Big(
    \frac{1}{2} a_\ka^\dagger A(\ka,\ka') a_{\ka'}^\dagger +
    a_{\ka'}^\dagger B(\ka,\ka') a_{\ka} \Big)+
    \int_{-\infty}^\infty d\ka \; V_2(\ka) a_\ka^\dagger + \frac{1}{2}V_{00}\,.
\end{align}
Here the matrices $A,B$ are given by
\begin{align}
A(\ka,\ka')=&\frac{\pi}{\N(\ka)}\delta(\ka+\ka')\,,\\
B(\ka,\ka')=&\frac{\ka \pi}{4}\coth(\frac{\ka \pi}{2})\delta(\ka-\ka')+
\frac{\ka+\ka'}{2}\delta'(\ka-\ka')\,.
\end{align}
The linear terms in the matter sectors are proportional to the
zero mode momentum,
\begin{equation}
V_1(\ka)=V_2(\ka) = -\frac{\pi}{2\sqrt{\N(\ka)}}
\tanh(\frac{\ka\pi}{4})
p_0\,,\\
\end{equation}
and for the bosonic ghost they depend on the ghost number and are given by
\begin{align}
\label{V1eq}
V_1(\ka) &= \p\frac{\pi}{2\sqrt{\N(\ka)}}\Big(
    3\coth(\frac{\ka\pi}{2})
    - q_0\tanh(\frac{\ka\pi}{4})\Big),\\
\label{V2eq}
V_2(\ka) &= \p\frac{\pi}{2\sqrt{\N(\ka)}}\Big(
    - 3\frac{1}{\sinh(\frac{\ka\pi}{2})}
    - q_0\tanh(\frac{\ka\pi}{4})\Big).
\end{align}
With these conventions $q_0$ refers to the ghost number of the ket
state.
One can also work in a more (anti)symmetric
convention,
\begin{align}
\tilde V_m(\ka)=-\p\frac{\pi}{2\sqrt{\N(\ka)}}\Big(
    (-1)^m\frac{3}{2}\coth(\frac{\ka\pi}{4})
    + \tilde q_0\tanh(\frac{\ka\pi}{4})\Big), \qquad
\tilde q_0 \equiv q_0-\frac{3}{2}\,.
\end{align}
In this convention $\tilde q_0$ is an anti-Hermitian operator.
This
is the usual convention-ambiguity of a linear dilaton theory.
We prefer to work with the conventions of~(\ref{V1eq},\ref{V2eq})
as we did in~\cite{Fuchs:2006an}, where we already illustrated
that the zero mode dependence of the algebra closes, so we just set $q_0=0$.
Also, in section~\ref{sec:V1V2V3} we calculate inner products with kets
with $q_0=0$, so again it is convenient to use these conventions.
Lastly, the constant terms are $V_{00}=p_0^2$ for the matter sector
and $V_{00}=q_0(q_0-3)$ for the ghost sector.

When we calculated the
commutation relation we found that the infinities cancel leaving an anomaly
for the finite part. The commutation relation that we found was,
\begin{align}
[\cL_{0,\text{total}},\cL_{0,\text{total}}^\dagger]=
\cL_{0,\text{total}}+\cL_{0,\text{total}}^\dagger+C\,,
\end{align}
where $C$ is a constant. The contribution to this constant comes from two
sources,
\begin{align}
&[\cL_{0,\text{quad}},\cL_{0,\text{quad}}^\dagger]= C_1+
  \text{operators}\,,\\
&[\cL_{0,\text{lin}},\cL_{0,\text{lin}}^\dagger]= C_2 +
  \text{zero mode terms}\,.
\end{align}
Here $\cL_{0,\text{quad}}^\dagger$ is the term containing
$a^\dagger A a^\dagger$, while $\cL_{0,\text{lin}}$ represent
the linear terms.
We already showed in~\cite{Fuchs:2006an} that the operator and zero
mode terms in the algebra work out correctly.
Here we only analyze the constants $C_1, C_2$ that come from the
central charge.

The total constant is
\begin{align}
C=27C_1 + C_2\,.
\end{align}

Since our regularization is state oriented, we need to clear up
how exactly we apply it for the algebra calculation.
One way to view it is as if we are calculating
\begin{equation}
\bra{0}\cL_{0,\text{total}}\cL_{0,\text{total}}^\dagger\ket{0} = C\,.
\end{equation}
Alternatively, we can use the generalization of our regularization
for operators, as described in the introduction.
Actually, we can use directly either~(\ref{VirasoroAlgebra})
or~(\ref{ScalarAlgebra}).
This is the case since the matrix $B$ is not involved in the calculation,
and therefore there are no extra normal ordering issues.

We can now evaluate
\begin{align}
C_2(\e)=-\Big(\frac{3\pi}{2}\Big)^2
\int_{-\infty}^\infty d\ka d\ka'
 \p \frac{\coth \big(\frac{\ka' \pi }{2}\big)}
         {\sinh \left(\frac{\ka \pi }{2}\right)}
\frac{\rho_\e(\ka,\ka')}{\sqrt{\N(\ka)\N(\ka')}}\,.
\end{align}
We can take care of the principal part prescriptions by
anti-symmetrizing~(\ref{rho}) with respect to $\ka\rightarrow -\ka$.
The anti-symmetrization with respect to $\ka'$ is then automatic.
Thus,
\begin{align}
\nonumber
C_2(\e)=-\Big(\frac{3\pi}{2}\Big)^2 &
\int_{-\infty}^\infty d\ka d\ka'
\frac{\coth \big(\frac{\ka' \pi }{2}\big)}
  {8\sinh^2 \left(\frac{\ka \pi }{2}\right)
   \sinh \left(\frac{\ka' \pi }{2}\right)}\cdot \\&
\label{C2}
\cdot \left( \frac{e^{i \frac{\ka-\ka'}{2} \log \e}}
       {B(\frac{i \ka'}{2},-\frac{i \ka}{2})}+
       \frac{e^{-i \frac{\ka-\ka'}{2} \log \e}}
       {B(\frac{i \ka}{2},-\frac{i \ka'}{2})}+
       \frac{e^{i \frac{\ka+\ka'}{2} \log \e}}
       {B(-\frac{i \ka'}{2},-\frac{i \ka}{2})}+
       \frac{e^{-i \frac{\ka+\ka'}{2} \log \e}}
       {B(\frac{i \ka}{2},\frac{i \ka'}{2})}
\right).
\end{align}
Following our general prescription, we can evaluate the $\ka'$
integral by dropping the second and third summands and evaluating
the residues of the two remaining summands at $\mbox{$\ka'=0,\pm \ka$}$.
The evaluation of the $\ka'$ residues leaves us with
\begin{align}
C_2(\e)=-\Big(\frac{3\pi}{2}\Big)^2 &
\int_{-\infty}^\infty d\ka
\frac{\frac{\ka \pi}{2} \coth \big(\frac{\ka \pi }{2}\big)-
  \cos \big(\frac{\ka \log \e}{2}\big)}
{\pi \sinh ^2\left(\frac{\ka \pi }{2}\right)}=\frac{9}{2}\big(\log \e+1\big).
\end{align}
This result exactly matches the one obtained in~\cite{Fuchs:2006an}.

For $C_1(\e)$ we have
\begin{align}
\nonumber
C_1(\e)&=\frac{\pi^2}{32} \int_{-\infty}^\infty d\ka d\ka'
\frac{1}
  {\sinh^2 \left(\frac{\ka \pi }{2}\right)
     \sinh^2 \left(\frac{\ka' \pi }{2}\right)}
\left( \frac{e^{i \ka_- \log \e}}
       {B(\frac{i \ka'}{2},-\frac{i \ka}{2})}+
       \frac{e^{-i \ka_- \log \e}}
       {B(\frac{i \ka}{2},-\frac{i \ka'}{2})}
\right)^2
\\&= \frac{\pi^2}{32} \int_{-\infty}^\infty d\ka d\ka'
\left(X_1+X_2+X_3 \right),
\end{align}
where we defined
\begin{align}
X_1=&
\frac{e^{i(\ka-\ka')\log\e}}
  {\sinh^2 (\frac{\ka \pi }{2})\sinh^2 (\frac{\ka' \pi }{2})
  B^2(\frac{i \ka'}{2},-\frac{i \ka}{2})},\\
X_2=&
\frac{e^{-i(\ka-\ka')\log\e}}
  {\sinh^2 (\frac{\ka \pi }{2})\sinh^2 (\frac{\ka' \pi }{2})
  B^2(\frac{i \ka}{2},-\frac{i \ka'}{2})},\\
X_3=&
\frac{\ka \ka' }
{\pi(\ka-\ka') \sinh\big(\frac{\ka-\ka'}{2} \pi \big)
  \sinh\big(\frac{\pi\ka}{2} \big)\sinh\big(\frac{\pi\ka'}{2} \big)}\,.
\end{align}
Now, $X_2$ does not contribute and $X_1$ contributes from the
residue at $\ka'=\ka$, which gives
\begin{align}
\label{C1Eps}
\int_{-\infty-i\delta}^{\infty-i\delta} d\ka' X_1\cong
-\frac{\ka^2\left(2\log\e+2\gamma+\psi(\frac{i\ka}{2})+
     \psi(-\frac{i\ka}{2})\right)}
 {\pi\sinh^2 \left(\frac{\ka \pi }{2}\right)}\,,
\end{align}
where in the r.h.s we have neglected an anti-symmetric integrable
imaginary part that would cancel out after the $\ka$ integration.
Next, we have to evaluate the integral of $X_3$. Here we have to sum
all residues, since this term is $\e$ independent.
Again we neglect the anti-symmetric part and get
\begin{align}
\label{C1noEps}
\int_{-\infty-i\delta}^{\infty-i\delta} d\ka' X_3\cong
\frac{\ka^2\left(2\gamma+\psi(\frac{i\ka}{2})+
     \psi(-\frac{i\ka}{2})-2\right)}
 {\pi\sinh^2 \left(\frac{\ka \pi }{2}\right)}\,.
\end{align}

Adding~(\ref{C1noEps}) and~(\ref{C1Eps}) we see that the terms
resembling $\rho_{\text{fin}}$ actually cancel out and we are left
with the simple integral
\begin{align}
-\frac{2}{\pi}\big(\log\e+1\big)\int_{-\infty}^{\infty} d\ka
\frac{\ka^2}{\sinh^2 \left(\frac{\ka \pi }{2}\right)}=
-\frac{16}{3\pi^2}\big(\log\e+1\big)\,.
\end{align}
Thus,
\begin{align}
C_1(\e)=-\frac{1}{6}\big(\log\e+1\big)\,.
\end{align}
$$\vphantom{the next page is horribly empty}$$
$$\vphantom{the next page is horribly empty}$$
This result differs from the expression in~\cite{Fuchs:2006an}.
We can now sum and get
\begin{align}
C(\e)
=27C_1(\e)+C_2(\e)=0\,,
\end{align}
the same expression as one gets using the Virasoro algebra with vanishing
central charge.

\subsection{Using Virasoro operators}
\label{sec:SchnablVir}

We can repeat the calculation using a Virasoro algebra with
non vanishing central charge,
\begin{align}
\nonumber
C_c(\zz)
=\sum_{n=1}^\infty \left(\frac{-2(-1)^n \zz^{2n}}{4n^2-1}\right)^2
    \frac{c}{12}\big((2n)^3-(2n)\big) = &
    \frac{c}{6}\left(\frac{1+\zz^2}{\zz^2}\tan^{-1}(\zz^2) - 1 \right)\\ =&
    -\frac{c}{6}\Big( \log \e+1+
    \cO(\e)
     \Big)\,.
\end{align}
This demonstrates that we have indeed managed to define a
consistent regularization scheme for the oscillator representation
and that this regularization is
universal.

As opposed to the continuous basis calculation where we only calculated
to the required order in $\e$, the calculation of $C_c$ using the Virasoro
generator is exact.
This raises the hope that one could generalize relations such as,
\begin{equation}
\label{nonLinLL}
e^{t(\cL_0+\cL_0^\dagger)}=
    \Bigl(\frac{2}{2-2t}\Big)^{\cL_0^\dagger}
    \Bigl(\frac{2}{2-2t}\Big)^{\cL_0}\,,
\end{equation}
to CFT's with a non vanishing central charge.
Turning on a central charge adds an infinite normalization to
this relation, which can be made finite using our regularization.
Getting this relation requires repeated use of Schnabl's algebra
and therefore depends on the exact form of the regularized algebra.
Unfortunately, $C_c$ is not enough, since the operators in the
algebra also get $\cO(\e)$ corrections.
We now verify that indeed such subleading terms emerge
and calculate them explicitly.

The geometric nature of our regularization allows for a straightforward
definition of the regularized Schnabl operators using
conformal transformations.
Let $u=\zz \xi$ and $v=-\zz \xi^{-1}$,
then by definition the regularized operators are given by
\begin{align}
\cL_0^\zz
=&\frac{1}{2\pi i}\oint du(1+u^2)\tan^{-1}(u)T(u)=
 \frac{1}{2\pi i \zz}\oint d\xi(1+(\zz\xi)^2)\tan^{-1}(\zz\xi)T(\xi)\,,\\
{\cL_0^\zz}
^\dag=&\frac{1}{2\pi i}\oint dv(1+v^2)\tan^{-1}(v)T(v)=
 \frac{1}{2\pi i \zz}\oint d\xi
    (\xi^2+\zz^2)\tan^{-1}\big(\frac{\zz}{\xi}\big)T(\xi)\,,
\end{align}
where in the last line the sign was reversed due to the orientation change.
Now,
\begin{align}
\nonumber
[\cL_0^\zz,{\cL_0^\zz}^\dag]&=
\frac{1}{2\pi i \zz^2}\oint d\xi \\
\nonumber
& \res_{\tilde \xi=\xi}\Big(
    (\xi^2+\zz^2)\tan^{-1}\big(\frac{\zz}{\xi}\big)
      (1+(\zz\tilde \xi)^2)\tan^{-1}(\zz\tilde \xi)
\big(\frac{2}{(\tilde \xi-\xi)^2}T(\xi)+\frac{1}{\tilde \xi-\xi}\partial_\xi T(\xi)\big)
\Big)\\
\nonumber
=&\cL_0^\zz+{\cL_0^\zz}^\dag
-\frac{2(1-s^4)}{2\pi i \zz^2}\oint d\xi
\xi \tan^{-1}(\zz\xi)\tan^{-1}\big(\frac{\zz}{\xi}\big)T(\xi)\\
\nonumber
=&\cL_0^\zz+{\cL^\zz}_0^\dag
-2(1-s^4)\sum_{k,m=0}^\infty
\frac{(-1)^{m+k}s^{2(m+k)}}{(2m+1)(2k+1)}L_{2(m-k)}\\
=&\cL_0^\zz+{\cL_0^\zz}^\dag
-(1-s^4)\bigg(\frac{1}{2} \Phi \big(\zz^4,2,\frac{1}{2}\big)L_0+\\
\nonumber
& \sum_{n=1}^\infty
(-1)^n \zz^{2 n} \Big(\frac{\tanh^{-1}(\zz^2)}{n \zz^2}-
   \frac{\,_2F_1(1,n+\frac{1}{2};n+\frac{3}{2};\zz^4)}{2n^2+n}\Big)(L_{2n}+L_{-2n})\bigg)
   \\ \nonumber
= &\cL_0^\zz+{\cL_0^\zz}^\dag
-\e\Big(\pi^2 L_0+2\sum_{n=1}^\infty
\frac{(-1)^n \left(2 H_{2 n-1}-H_{n-1}\right)}{n}
(L_{2n}+L_{-2n})\Big)+\cO(\e^2 \log \e).
\end{align}
Here, $\Phi$ is the Lerch transcendent, $_2F_1$ is the hypergeometric function and
$H$ is the harmonic number.
We see that while our regularization obeys Schnabl's algebra in the $\e\rightarrow 0$
limit, there are $\cO(\e)$ corrections to it.

\subsection{A generalization of the regularization}
\label{sec:SchnablAlt}

The fact that our regularization respects Schnabl's algebra only up to
$\cO(\e)$ means that we cannot use it to regularize operators that are
general nonlinear functions of $\cL_0,\cL_0^\dag$.
It would have been desirable to find a regularization that is
adequate also for this case, especially since wedge states and string
vertices, which we consider in the next section, can be represented in such
a way.
Ideally we would like to have a regularization that respects exactly the full
Virasoro algebra. This, however, seems too optimistic.
The reason is that the Virasoro algebra encodes all the geometric data
while the origin of singularities, that one wants to resolve, is also
geometrical.
In particular, one can easily see that a universal regularization
necessarily modifies the Virasoro algebra.
Still, one can hope to find a regularization that would leave Schnabl's
algebra intact, without the $\cO(\e)$ corrections that we
evaluated
in section~\ref{sec:SchnablVir}.
We are not sure if such a regularization exists. However, if there is such
a regularization it can be used to regularize expressions such
as~(\ref{nonLinLL}) and then it can be considered as a regularization of
operators and not only states.

In the continuous basis a natural ansatz for a generalization of
our regularization is
\begin{align}
\rho_\e(\ka,\ka')=\frac{1}{2\pi i}
     \Big(\frac{e^{-i \frac{\ka-\ka'}{2} \log \e}}
       {\ka-\ka'}f(\ka,\ka')-\frac{e^{i \frac{\ka-\ka'}{2} \log \e}}
       {\ka-\ka'}f^*(\ka,\ka')\Big).
\end{align}
The origin of the regularization can be
geometric, but is not necessarily so.
In order to for this expression to be a resolved delta function
we should demand $f(\ka,\ka)=1$.
We would also like the regularization to be symmetric, that is
\begin{align}
\rho_\e(\ka,\ka')=\rho_\e(\ka',\ka)\,.
\end{align}
Together with a reality requirement this restricts the $f(\ka,\ka')$
to take the form
\begin{align}
& f(\ka,\ka')=f_s+i f_a\,,\\
\nonumber
& f_s=1+k_-^2 f_2(\ka_+)+k_-^4 f_4(\ka_+)+...\,,\qquad
f_a=k_- f_1(\ka_+)+k_-^3 f_3(\ka_+)+...\,.
\end{align}
It is important to stress that while all these regularizations refine
the delta function, which seems as the source of divergences in the
continuous basis, they are not necessarily consistent, since a priori they
are not related to the symmetries of the theory.

The simplest generalization is given by $f(\ka,\ka')=1$,
but we have to check its consistency.
For this case the multiplication of regulators gives
\begin{align}
\rho_\e\rho_{\e'}=\rho_{\min(\e,\e')}\,.
\end{align}
This result, unlike~(\ref{rhorho}) is exact (no $\cO(\e)$ corrections).
This implies that these regulating matrices are projectors
and as such can only have zeros and ones as their eigenvalues.
It is easy to diagonalize these projectors. For that we
note that since they are functions only of $\ka-\ka'$ the multiplication of
two such operators is just a convolution. Going to the variable $x$, which is
the Fourier transform of
$\ka-\ka'$ turns the convolution into a regular multiplication. Then
\begin{align}
\rho_\e(x)=\theta (x-\log \e)-\theta (x+\log\e)\,.
\end{align}
This regularization amounts to truncating the frequencies
to the range $\log \e <x< -\log\e$.
We repeated the calculations of section~\ref{sec:SchnablOs} using
this regularization to check its consistency.
The results are that $C_2(\e)=\frac{9}{2}(\log \e+1)$,
as in section~\ref{sec:SchnablOs} and in the oscillator
level truncation evaluation~\cite{Fuchs:2006an}.
The quadratic term, however, gives a different and thus inconsistent, result
$27C_1(\e)= -\frac{9}{2}(\log \e+\log 2-\frac{1}{4})$.
This result reminds us of the discussion at the end of~\cite{Okuyama:2002tw}.
Once again we see that it is not enough to resolve the singularity,
but one should do it while respecting the symmetries.
In~\cite{Okuyama:2002tw} the singularity originated from divergences at $\ka=0$
in the continuous basis and the symmetry was a specific algebra obeyed by the
matrices defining the string vertex.
Here, we have to resolve the delta function, since otherwise we get $\delta^2$
terms and the relevant symmetry is universality.

We now want to see if there are other members of this family for which
the constant in the commutation relation vanishes.
The evaluation of $C_2(\e)$ is straightforward and gives
\begin{align}
\label{C2alt}
C_2(\e)=\frac{9}{2}\big(\log \e+1+2f_1(0)\big)\,.
\end{align}
For $C_1(\e)$ the equations are not as simple. The result is
\begin{align}
\label{C1alt}
27C_1(\e)=-\frac{9}{2}(\log \e+J_1+J_2)\,,
\end{align}
where
\begin{align}
J_1=\frac{3\pi}{8}\int_{-\infty}^\infty
 \frac{\ka^2 f_1(\ka)}{\sinh^2(\frac{\ka \pi}{2})}d\ka\,,\qquad
J_2=-\frac{3}{32}\int_{-\infty}^\infty
 \frac{\ka \ka' |f(\ka,\ka')|^2}
  {\ka_-^2\sinh(\frac{\ka \pi}{2})\sinh(\frac{\ka' \pi}{2})}d\ka d\ka'\,.
\end{align}
Note that the $\ka'$ integration contour of $J_2$ lies below that of $\ka$ and
that $J_2$ depends on all the $f_j$'s.
We see that the infinite part always cancels.
The final result would be consistent provided that
\begin{align}
J_1+J_2=1+2f_1(0)\,.
\end{align}
There are infinitely many choices for such regularizations. The reason is that
for a given $f$ it is always possible to change $f_1$ at the vicinity
of $\ka_+=0$,  without modifying $J_1$ and $J_2$ by much.

\section{Wedge states and the descent relations of string vertices}
\label{sec:V1V2V3}

We want to show that the descent relations
\begin{equation}
\braket{V_1}{V_{N}}=\ket{V_{N-1}},
\end{equation}
hold without a need for any extra normalization.
To simplify the above relation we contract it with
$N-1$ vacuum states, of which $N-2$ have ghost number zero
and one vacuum state is taken with ghost number
three, to saturate the ghost number.
This leaves us with a relation among wedge states
\begin{equation}
\braket{1}{N}=1\,,
\end{equation}
where one state has ghost number zero
and the other has ghost number three.
In what follows we consider the more general case of wedge state
inner product $\braket{n_1}{n_2}$, where $n_i$ are not necessarily
integers.
This should also equal unity, as should be the case for inner product of
any two surface states.
Also, note that the location of the ghost insertion is not
important, as can be deduced from the CFT expression, where the two
wedge states form a half infinite cylinder of a given circumference
and the $c^3$ insertion can be moved around the boundary of this cylinder.

We define the wedge states (before regularization) by
\begin{align}
\label{braAB}
\bra{n}&=\bra{0}e^{\log\frac{2}{n}\cL_0}=\bra{0}
e^{\log\frac{2}{n}(\frac{1}{2}a A a+a^\dag B a+
 a V_1)}\,,\\
\label{ketAB}
\ket{n}&=e^{\log\frac{2}{n}\cL_0^\dag}\ket{0}=
e^{\log\frac{2}{n}(\frac{1}{2}a^\dag A a^\dag+a^\dag B^T a+
 a^\dag V_2)}\ket{0}.
\end{align}
This convention guaranties that the overlap of all wedge states with
the vacuum (wedge state $n=2$) is unity for any central charge.
It is the standard convention used in the literature.
Then, in order to evaluate the inner product it is convenient
to write
\begin{align}
\label{alphabeta}
\bra{n}=\bra{0}e^{ \frac{1}{2}a \al(n) a+
 a \beta_1(n) }\,,\qquad\qquad
\ket{n}=e^{ \frac{1}{2}a^\dag \al(n) a^\dag+
 a^\dag \beta_2(n)
}\ket{0}.
\end{align}
In order to find the matrix $\al$ and the vectors $\beta_m$
we define
\begin{align}
u\equiv \log\frac{2}{n}
\end{align}
and derive~(\ref{braAB},\ref{ketAB},\ref{alphabeta}) with respect to
$u$.
Using the commutation relations we arrive at the differential equations
\begin{align}
\al'(u) &= A + B^T\al(u) + \al(u)B\,,\\
\beta'_m(u) &= V_m + B^T\beta_m(u)\,,
\end{align}
whose solution is
\begin{align}
\label{alpha}
\al(u) &=
\sum_{n=1}^\infty \sum_{k=0}^{n-1}u^n
   \frac{(n-1)!}{(n-1-k)!k!}{B^T}^k A B^{n-1-k}\,, \\
\label{beta}
\beta_m(u) &=
\sum_{n=1}^\infty \frac{u^n}{n!} {B^T}^{n-1} V_m\,.
\end{align}

Regularization amounts to replacing
\begin{align}
A \rightarrow A^\zz = \rho_\zz^\frac{1}{2} A \rho_\zz^\frac{1}{2}\,,\qquad
B \rightarrow B^\zz = \rho_\zz^{-\frac{1}{2}} B \rho_\zz^\frac{1}{2}\,,\qquad
V_m \rightarrow V_m^\zz = \rho_\zz^\frac{1}{2} V_m\,.
\end{align}
Notice the negative power of $\rho_\zz$
in the definition of $B^\zz$. It is a result of the fact
that one index of the matrix contracts a creation operator while the
other index contracts an annihilation
operator (recall that our regularization is state oriented).
One could have been worried about the negative powers of the regulator,
but in fact this feature simplifies the rest of the regularization
and only positive powers of $\rho_\zz$ appear in the final expression
\begin{align}
\al^\zz &= \rho_\zz^\frac{1}{2} \al \rho_\zz^\frac{1}{2}\,,\\
\beta_m^\zz &= \rho_\zz^\frac{1}{2}\beta_m\,.
\end{align}
Actually, this simplification was bound to occur due to the nature
of the regularization, where the power of $\zz$ counts the
conformal weight of the operators in the state.

In subsection~\ref{sec:cont} we evaluate these descent relations in the
continuous basis, where the wedge states are diagonal\footnote{The
only surface states with simple representations in
the continuous basis are the wedge states and the
butterflies~\cite{Uhlmann:2004mv,Fuchs:2004xj}.}.
Then, in subsection~\ref{sec:disc} we study
them also in the discrete basis.

\subsection{Continuous basis calculations}
\label{sec:cont}

Before regularization we can use the differential equations
to get a closed form for their solutions,
\begin{align}
\label{alphaTildeSol}
\alpha(n) &=\frac{\sinh(\frac{(2-n)\ka\pi}{4})}
  {\sinh\big(\frac{n\ka\pi}{4}\big)}\delta(\ka+\ka')\,,\\
\beta_{\text{mat}}(n) &=\p\frac{\sinh(\frac{\ka \pi}{2})}{\ka \sqrt{\N(\ka)}}
\Big( \tanh\big(\frac{n \ka \pi}{8}\big) -
      \tanh\big(\frac{\ka \pi}{4}\big) \Big)p_0,\\
\label{beta1TildeSol}
\beta_1(n) &=\p\frac{3}{\ka \sqrt{\N(\ka)}}
\Big(\frac{\sinh(\frac{\ka \pi}{2})}
  {\sinh(\frac{n\ka \pi}{4})}-1 \Big),\\
\label{beta2TildeSol}
\beta_2(n) &=\p\frac{3}{\ka\sqrt{\N(\ka)}}
   \Big(\cosh(\frac{\ka \pi}{2})-\coth (\frac{n\ka \pi}{4})
 \sinh(\frac{\ka \pi }{2}) \Big).
\end{align}
The expression for $\al(n)$ is simply the unregularized expression
for the wedge states found already in~\cite{Furuuchi:2001df}.
In what follows $\beta_{\text{mat}}$ will not be used,
it is only given for completeness.

To regularize these expressions we have to add the $\zz$ superscript to
$\al(n),\beta_m(n)$ or equivalently, to insert $\rho^\zz$
whenever two indices are being contracted.  This $\zz$ dependence will be
implicit in the rest of this subsection
for notational simplicity. All expressions should
be understood as regularized.
We now use the formulas of~\cite{Kostelecky:2000hz} to write the
normalization
\begin{align}
\label{KPexp}
\braket{n_1}{n_2}=\exp\Big(-&\frac{27}{2}
   \tr \log\big(1-\al_1\al_2\big)+\\&
\nonumber
\beta_2\big(1-\al_1\al_2\big)^{-1}\beta_1+
\frac{1}{2}\beta_2\big(1-\al_1\al_2\big)^{-1}\al_1\beta_2+
\frac{1}{2}\beta_1\big(1-\al_2\al_1\big)^{-1}\al_2\beta_1\Big),
\end{align}
where $\beta_m\equiv \beta_m(n_m)$ and $\al_m\equiv \al(n_m)$.

We have not managed to evaluate this expression analytically as a
function of $u_m$. However, in the limit $\e\rightarrow 0$
the result should tend to unity 
at least $\forall u_m<\log2$.
Thus, we can
expand the argument inside the exponent in a double Taylor series
in $u_m$ and all the coefficients should cancel between the first
and second lines of~(\ref{KPexp}).
A useful expression in these calculations is
\begin{align}
\label{rhoArho}
(\rho^\zz A\rho^\zz)
(\ka,\ka')=\frac{-1}{2\sqrt{\N(\ka)\N(\ka')} (\ka+\ka')}
\Big(\frac{e^{i \frac{\ka+\ka'}{2} \log\e}}{\ka+\ka'+2i}+
    \frac{e^{-i \frac{\ka+\ka'}{2} \log\e}}{\ka+\ka'-2i}\Big)\,.
\end{align}
The calculations are somewhat lengthy and not too illuminating.
The result is
\begin{align}
&\nonumber
\beta_2 \big(1 - \al_1\al_2\big)^{-1}\beta_1+
\frac{1}{2}\beta_2\big(1-\al_1\al_2\big)^{-1}\al_1\beta_2+
\frac{1}{2}\beta_1\big(1-\al_2\al_1\big)^{-1}\al_2\beta_1 = \\
&\nonumber
\frac{27}{2}\tr \log \big(1-\al_1\al_2\big) = \\
&\nonumber
\qquad
  u_1 \frac{9}{4}(\log\e+1)(e^{2u_2}-1)
  +u_1^2 \Big (\frac{63\log\e+27+6\pi^2}{8} u_2^2+
 \frac{69\log\e+30+3\pi^2+36\zeta(3)}{8} u_2^3\\
&\label{uiuj}
\qquad\quad
  +\frac{1095 \log \e+495+40 \pi ^2+8 \pi^4}{160}u_2^4\\
&\nonumber
\qquad\quad
  +\frac{681\log\e+318+15 \pi ^2+180 \zeta (3)-4 \pi ^4+720 \zeta (5)}{160}
     u_2^5
+\cO(u_2^6)\Big )+\cO(u_1^3)\,.
\end{align}
Part of the (double) integrals
leading to the constant contributions in the last two lines were calculated
numerically for the trace calculation.
The final numerical results agree with the one stated
up to at least six significant digits.
All the expressions related to the second line of~(\ref{KPexp})
were calculated analytically. In order to perform the integrals one should
keep in mind the principal part in the definition
of~(\ref{beta1TildeSol},\ref{beta2TildeSol}), which is taken care of by
antisymmetrizing the $\rho$ that acts on it with respect to the relevant
variable. Actually, antisymmetrization with respect to any of the two
variables gives the same result. One should also be careful with contour
deformations. In the calculations triple integrals appear. In order to perform
these integrals all three contours should be deformed. The relative
``heights'' of the three contours in the complex plane effects the evaluation.
The final result is of course independent of the chosen deformation.
Also, while the coefficients of say $u_1^2 u_2^3$ and $u_1^3 u_2^2$ coming
from~(\ref{beta1TildeSol},\ref{beta2TildeSol}) are not a priori equal,
it is clear that they must be the same, since they both equal the
coefficient of the trace part. Practically, instead of calculating both,
one can evaluate one of them and their difference, which can be shown to
vanish relatively easy. Indeed we have checked these coefficients
($u_1^n u_2$ and $u_1^n u_2^2$) and the results agree.

We note that each one of the coefficients of the
$u_1^2 u_2^n$ terms contains in addition to the $\log \e$ factor also a
rational term and a sum of rational coefficients times $\zeta(k)$ for
$2\leq k\leq n$. In order to decipher the meaning of this we switch from
the $u$ variable which we used in order to find $\al$ and
$\beta_m$
to
\begin{align}
t=-\frac{n-2}{2}=1-e^{-u}\,.
\end{align}
In these coordinates our results read
\begin{align}
\frac{27}{2}& \tr \log \big(1-\al_1\al_2\big)=
t_1 \frac{9}{4}(\log\e+1)\frac{(2-t_2) t_2}{(1-t_2)^2}+\\& \frac{9}{8}t_1^2
\sum_n \Big((n+1)(n+2)(\log\e+1)
\nonumber
-\frac{(n-1) (n+2)^2}{2 n}+2\sum_{k=2}^n (n-k+2)\zeta(k)
\Big) t_2^n+\cO(t_1^3).
\end{align}
We have calculated these coefficients for $n=1..5$. However, if we can trust
it as a general formula then we can sum the coefficients of $t_1^2$ to get
\begin{align}
\label{t2Co}
\frac{9 \left(3-3 t_2+t^2_2\right)t_2}{4 (1-t_2)^3}(\log\e+1)
 -\frac{9 (2-t_2) t_2}{8 (1-t_2)^3}-\frac{9}{4} \log (1-t_2)
 -\frac{9(2-t_2) t_2 \left(\psi (1-t_2)+\gamma \right)}{4 (1-t_2)^2}\,.
\end{align}
We have managed to evaluate the infinite parts analytically and indeed
they match the above expression both for the trace and for the other terms.
We evaluated the finite parts numerically.
These parts match each other and match~(\ref{t2Co})
to more than seven significant digits up to $t_2\approx .7$, where numerical
integration starts to lose its precision due to
the singularity at $t_2=1$. Actually, there should be a
singularity already at $t_i=\frac{1}{2}$ that corresponds to the identity.
This singularity is invisible at this order of calculation. 

To summarize we can write
\begin{align}
\braket{n_1}{n_2}=1+\cO(t_1^3 t_2^3)\,.
\end{align}
In all former evaluations of these expressions anomalies occurred
for all the non-trivial cases, that is, they were present in all orders
of the double Taylor series.
We thus view our result as a strong indication that as long as the
exact expression for~(\ref{uiuj})
is well defined, that is as long as $1\leq n_i\leq\infty$,
the inner product is exactly unity as it should be.
This is a strong argument in favor of our regularization.
In fact, we believe that this would be the result of using our
regularization with any well defined surface states.

\subsection{Discrete basis calculations}
\label{sec:disc}

We start by calculating $\braket{1}{3}$ numerically.
Using~(\ref{Squeezed}) we can simply use the known expressions
for the wedge states $\bra{1},\ket{3}$ and just insert the regularizing
factors accordingly.
There are several checks that we can do numerically.

First, we check that taking the matter and ghost sector separately
one gets finite results for every $\zz<1$. The numerical calculation is
only approximate since we are taking finite matrices.
For small values of $\zz$ the result converges very quickly to a constant.

Next, we check the dependence of this constant on $\e=1-\zz$ to see
that we get the $\log\e+1$ behaviour. This calculation can be a bit tricky
since for large values of $\e$ there are $\cO(\e)$ correction while
for small values of $\e$ we need large matrices for the calculation to
converge.
Luckily, current computer resources are strong enough to leave us with
a large enough window in the range $2^{-7}<\e<2^{-4}$ using
1000 by 1000 matrices to verify the analytic computation.
 
Lastly, we can compute the total normalization.
Here the computation is easier, since the normalization should
equal unity for every value of $\zz<1$. Still, we try to approach $s=1$
to demonstrate that the radius of convergence is indeed unity.
These results are summarized in
figures~\ref{fig:numerical},\ref{fig:numericalExp}.
\FIGURE{
\epsfig{figure=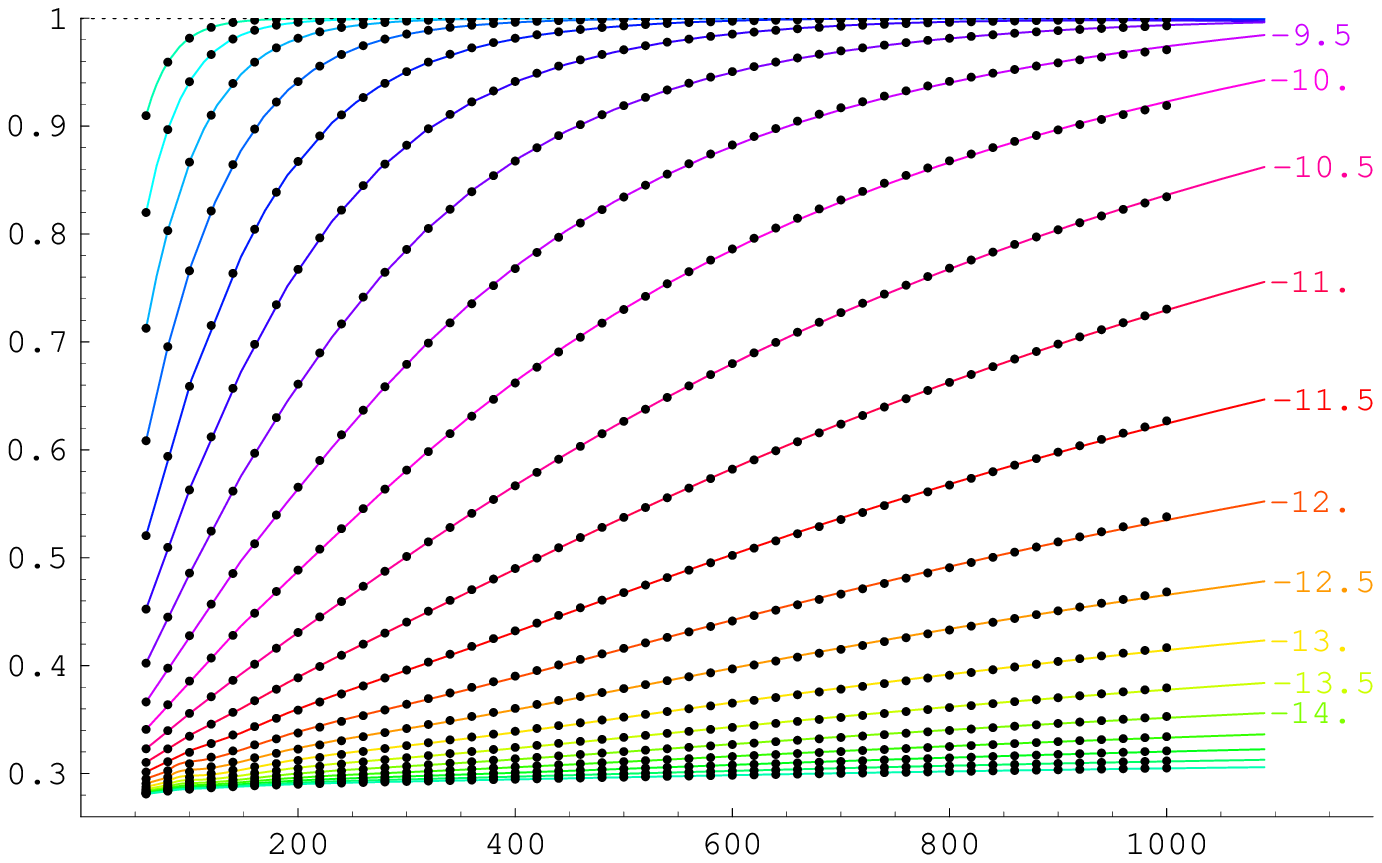}
\label{fig:numerical}
\caption{This plot describes the numerical calculation
of the $\gamma_{1,3}$ normalization.
The horizontal axis is the size of the matrix used
and the vertical axis is the normalization.
The different graphs are for different values of $\e$ between
$2^{-6}$ and $2^{-16}$, where some of the intermediate values
are labeled.
The dots are the calculated values and the lines are just an interpolation
we used to guide the eye.
For ``large'' values $(\e=2^{-6}\sim0.016)$ the results converge
to 1 very rapidly.
For extremely small values of $\e$ we will have to work with larger
and large matrices, but the pattern is clear -- eventually the
result converges to unity.
Interestingly, for $\e=0$, even if we take infinite matrices the result
will not converge to 1, but to 0.29.
}
}
\FIGURE{
\epsfig{figure=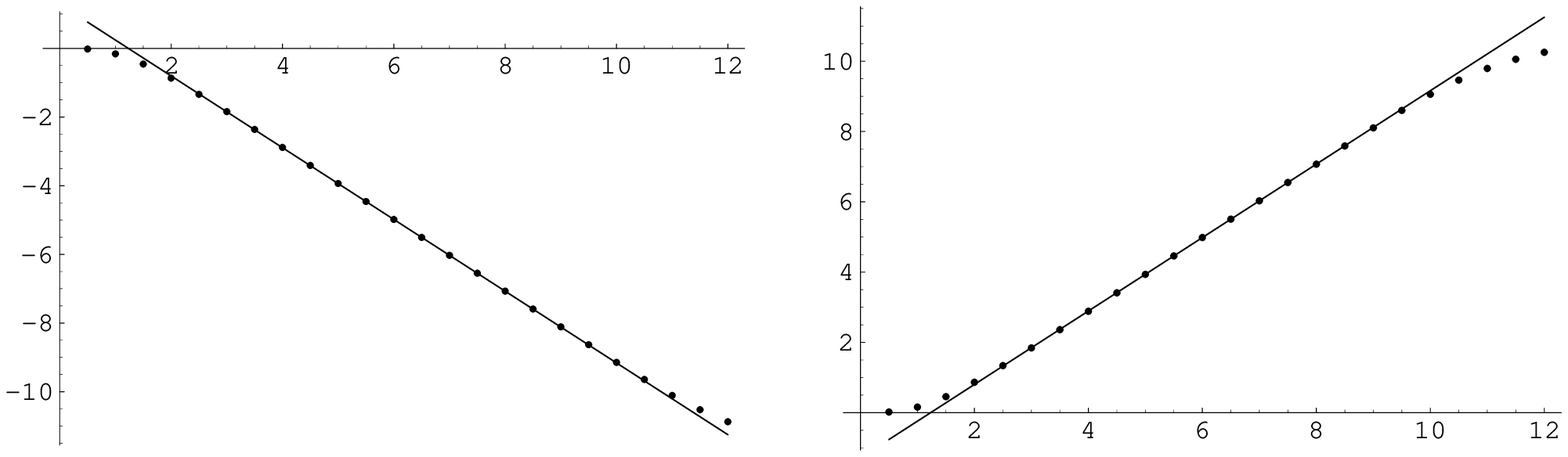,width=15cm}
\label{fig:numericalExp}
\caption{The left plot describes the normalization of the exponential term.
The $x$-axis is the $\log$ of the regularization parameter such that
$\e=0.5^x$. The $y$-axis is the argument of the exponent.
Therefore, the apparent linear behaviour corresponds to a $\log\e$
dependence of the normalization.
After ignoring edge points, a linear fit gives $y=-1.29-1.51\log\e$.
The result from continuous basis calculations is that
the coefficient of $\log\e$ is $-3/2$.
The deviation of the data points from the fit for small $x$ (large $\e$)
is a result of ignoring $\cO(\e)$ corrections.
The deviation of the data points for large $x$ is due to the fact that we
are using finite size matrices (1000 by 1000).
Up to a total sign the results for the determinant term (right plot)
are the same up to the prescribed accuracy.
The only qualitative difference is that the deviation for large $x$
starts a bit earlier for the determinant term.
This difference is the source of the anomaly seen in the oscillator level
truncation.
}
}

Another interesting case is the sliver normalization
$\braket{\infty}{\infty}$.
It is of interest because of the extensive use of the sliver
state in the literature and because the sliver seems more singular
than most other wedge
states~\cite{Rastelli:2001hh,Moore:2001fg,Gaiotto:2002kf}.
We did not try to optimize the numeric calculation and therefore
it was limited to 30 by 30 matrices.
Still this was enough to demonstrate that the normalization is unity
for $\zz\lesssim 0.9$.

We also computed the overlap of two identity states.
This calculation is much more singular since the surface state
touches the coordinate patch not only at the mid-point, but also at the entire
boundary. Truncating the matrix of the squeezed state also does not
help since it gives singular results for every matrix size.
Still, using our regularization we can get exact analytical results.
For $D$ scalar fields we get
\begin{equation}
\log\det(1-C^\zz C^\zz)^{-D/2}=-\frac{D}{2}\tr\log(1-C^\zz C^\zz)=
    -\frac{D}{2}\sum_{n=1}^\infty \log(1-\zz^{4n})\,.
\end{equation}
Showing that this expression has radius of convergence of unity proves
that our regularization gives a well defined value
to the $\braket{1}{1}$ overlap.
We used the scalar field for the calculation, but since the regularization
of the scalar field is equivalent to regularizing the Virasoro operators,
the above expression is universal after replacing the number of
scalar fields $D$ with their central charge $c$.

We are therefore guaranteed to get $\braket{1}{1}=1$ for a system with
central charge 0.
We still calculated this explicitly for the system of 26 scalar fields
and bosonized ghost. This demonstrates that the regularization works
in a highly non trivial way.
The expression for the log of the normalization is
\begin{equation}
f(x)\equiv \sum_{n=1}^\infty \bigg(-\frac{27}{2}\log(1-x^{n})+
    \frac{9}{2n}\frac{x^{n}}{1-x^{2n}} \Big(
    \big(2(-1)^n-1\big)-\frac{1}{2}x^{n}-
     \frac{1}{2}\big((-1)^n-2\big)^2 x^{n}
    \Big)\bigg),
\end{equation}
where we define $x\equiv s^4$.
To evaluate the expression we expand $(1-x^{2n})^{-1}$ in its Taylor
series as $\sum_{m=1}^\infty x^{2n(m-1)}$.
We now have a double sum term which we sum over $n$.
Renaming $m\rightarrow n$ we get
\begin{equation}
f(x)=\frac{9}{2}\log\prod_{n=1}^\infty \left(
   \frac{\left(1+x^n\right)^3 \left(1-x^{2n-1}\right)}
  {\left(1+x^{2 n}\right)^2 \left(1+x^{2n-1}\right)^2}\right)
   =\frac{9}{2}\log\prod_{n=1}^{\infty}(1+x^n)(1-x^{2n-1})\,.
\end{equation}
To see that this expression vanishes we use the identity
$(1+x^{2n-1})(1-x^{2n-1})=1-x^{2(2n-1)}$.
This leaves us with only even powers of $x$ and gives the relation
$f(x)=f(x^2)$, which means that $f(x)$ is a constant and this
constant is obviously zero at $x=0$.

\section{Conclusions}
\label{sec:conc}

In this paper we regularized the string field states by
a conformal transformation that shrinks them.
An important feature of this regularization is that it is independent
of the choice of CFT.
Level truncation regularization also shares this feature.
The added value of our regularization is analyticity with respect to
the shrink parameter $\zz$.
This allows us to study singularities that level truncation
cannot see, as was demonstrated with the butterfly state.

Our strategy is to turn on the regulator. Then we can consider
each sector of the CFT separately.
Without the regulator the calculations in each sector would 
diverge because of the non-zero central charge.
These are the infinities that we regularize.
After adding up the results from the different sectors,
the regularization is removed giving a consistent result.

Moreover, calculations in our regularization can be much simpler and
analytic calculations can be taken much further.
Specifically, we can apply our regularization to the continuous basis.
All previous calculations in the continuous basis used oscillator
level truncation, which is not universal, giving wrong results.

In the continuous basis we managed to calculate the normalization of the
commutation relation of Schnabl's operators and to check, to some extent,
the descent relations of the vertices.
It would be interesting to try and make an analytical calculation
for all the descent relations.
We did do some numerical calculations. Since it is impossible to use
infinite matrices, the calculation is again CFT dependent.
Still, we could take matrices up to 1000 by 1000 in size,
which can be considered infinite for $\zz\lesssim 0.99$.
Interestingly, for the most singular calculation, the calculation
of the normalization of the identity state, we
did manage to perform a fully analytic calculation.

For all singular calculations $\zz$ has
radius of convergence of unity.
One could ask what happens at $\zz=1$.
It seems that there could be a phase transition at this point,
which adds non trivial constants to the descent relations.
In~\cite{Belov:2003qt} it was suggested that indeed such
constants exist and that they can be explained by partition function
calculations, yet no one has performed these calculations.
In our regularization, this question is
irrelevant, since we avoid the need to deal with surfaces
with conical singularities.
All calculations are consistent in the limit $\zz\rightarrow 1$,
so we can define the value at $\zz=1$ to be the value of this limit.

Our regularization works for states and vertices.
Other string field theory calculations
would also require the use of the propagator.
There does not appear to be a straightforward generalization of our
regularization to the propagator operator.
We can hope that either such regularization will not be needed,
or that the intuitive geometric picture of our regularization
would allow for a simple generalization of our work to this case.

We could also think of generalizing our regularization by
replacing $s^{L_0}$ with other universal expressions.
One asked-for option is to use $s^{\cL_0}$ as the regulator.
With this choice the calculation of descent relations is much
simpler.
However, one can easily check that
the singularity in the descent relations is not removed.
So this operator cannot be used for regularization.
This gives us some intuitive understanding about the source of the divergence.
The conformal map related to $s^{\cL_0}$ shrinks the coordinate
patch everywhere except
at the mid-point, demonstrating that the mid-point is
``the source of all evil''.

Finally, we note that our work is relevant also to other,
non-polynomial string field theories. There has been recently an advance
in the field, with the definition of the heterotic string field
theory~\cite{Okawa:2004ii,Berkovits:2004xh} and the new results
from closed string field theory~\cite{Yang:2005rx,Moeller:2006cv}.
These theories are usually dealt with using universal expressions.
However, if one would attempt an oscillator approach, then relative
normalization coefficients would be even more important than in the case
of the cubic theory, where non-trivial factors can in principal be absorbed
into the string field.

\section*{Acknowledgments}

We would like to thank Dima Belov, Debashis Ghoshal, Yaron Oz,
Rob Potting, Cobi Sonnenschein and Stefan Theisen for discussions.
M.~K. would like to thank Tel-Aviv University for hospitality during
part of this project.
The work of M.~K. is supported by a Minerva fellowship.
The work of E.~F. is supported by the German-Israeli foundation for
scientific research and by a grant of the German-Israeli Project Cooperation -
DIP Program (DIP H.52).

\bibliography{FK}

\end{document}